\documentclass[aps,prl,twocolumn,floats]{revtex4}
\usepackage{color,ulem,verbatim,float,wrapfig}
\usepackage{physics}
\usepackage{amssymb,amsbsy,amsmath,mathrsfs}
\usepackage{appendix}
\usepackage{graphicx}
\usepackage{times}
\usepackage{color}
\usepackage{subfigure}
\usepackage{braket}
\usepackage{bbold}
\usepackage{commath}
\usepackage{setspace}
\usepackage{bm} 
\newcommand\bea{\begin{eqnarray}}
\newcommand\eea{\end{eqnarray}}
\newcommand\beq{\begin{equation}}
\newcommand\eeq{\end{equation}}

\newcommand{\si}{\sigma}

\newcommand{\ra}{\rangle}

\usepackage[colorlinks = true,
linkcolor = magenta,
urlcolor = magenta,
citecolor = magenta,
anchorcolor = magenta]{hyperref}
\usepackage{soul}
\usepackage[dvipsnames]{xcolor}
\begin{document}

\title{ Mechanisms of scrambling and unscrambling of quantum information in the ground state in spin chains: 
domain-walls, spin-flips and scattering phase shifts}
\author{Samudra Sur$^1$ and Diptiman Sen$^{1,2}$}
\affiliation{$^1$Center for High Energy Physics, Indian Institute of Science, Bengaluru 560012, India \\
$^2$Department of Physics, Indian Institute of Science, Bengaluru 560012, India}

\begin{abstract}
The spatiotemporal evolution of the out-of-time-order correlator (OTOC) measures the propagation and scrambling of local quantum information. For the transverse field 
Ising model with open boundaries, the local operator $\sigma^{x}$ shows an 
interesting picture of the ground state OTOC where the local information 
gets scrambled 
throughout the entire system and, more strikingly, starts `unscrambling' upon reflection at the other end. Earlier discussions of OTOCs did not
explain the physical processes responsible for such scrambling and
unscrambling of information. We explicitly show that in the paramagnetic phase, the scrambling and unscrambling is due to the scattering of a pair of low-energy spin-flip excitations, even in the presence of small interactions. In the ferromagnetic phase the same phenomena are explained by the motion of a domain-wall excitation. Thus, in different limits of the system parameters, we have provided a simple and almost complete understanding of the space-time pictures of the OTOCs, including the unscrambling, in terms of the low-energy excitations like one and two spin-flips or a single domain wall.

\end{abstract}

\maketitle

\paragraph{Introduction.}
\label{sec1}

The spreading of local quantum information in a quantum many-body system, known as scrambling~\cite{Roberts,hosur2016,swingle2017,patel2017quantum1,campisi2017thermodynamics,iyoda2018,pappalardi2018,klug2018,odell2019,alavirad2019,sahu2019scrambling,sur2022information}, is generally described by the growth of a local operator under time 
evolution. The front of the operator (or the light-cone) is known to spread ballistically~\cite{Luitz,bohrdt2017,khemani2018,rakovszky2018,von2018} in lattice 
systems with a bound on the speed of propagation called the Lieb-Robinson velocity~\cite{lieb1972finite}. Quantitatively, scrambling is measured by the 
out-of-time-order correlator (OTOC) \cite{larkin1969,shenker2014,maldacena2016,dora2017out,shen2017out,heyl2018detecting,bao2020out,mcginley2019slow,riddell2019out,dag2019detection,zamani2022out,shukla2021out, suchsland2022},
$F(l,t) = \braket{W_{l}(t) V_{0} W_{l}(t) V_{0}}$ or more precisely the real part of $F(l,t)$, where the local operators $V$ and $W$ are at the 
positions $0$ and $l$, and the expectation is taken in a suitable equilibrium ensemble. Recently, it has been found that in non-interacting spin-$1/2$ chains, the OTOCs for 
certain 
operators show scrambling of quantum information inside the light-cone (marked by the value of $F(l,t)$ deviating considerably from $1$), while for other operators they do
not~\cite{lin2018out,sur2022effect}. This phenomenon was attributed to the locality or non-locality of those operators in the Jordan-Wigner (JW) fermionic picture, but without a
quantitative explanation. So far, the approach to understand scrambling through OTOCs has been attempted mainly by studying the Heisenberg time-evolution of the local operator 
using the Baker-Campbell-Hausdorff commutator expansion. In this context, several works\cite{von2018,khemani2018,rakovszky2018,gopalakrishnan2018,nahum2018,agarwal2019,
lopez2021,zhou2023} in recent times have described the operator spreading in one
dimension in terms of a coarse-grained hydrodynamic picture of quasiparticle diffusion. 
While these explain the ballistic spreading and diffusive broadening of the
operator front, an exact mechanism for scrambling in terms of the excitations of a 
system has been elusive so far.

It has been reported~\cite{sur2022effect} in a recent study that the initial local operator starting from one end of the systems gets scrambled under time evolution, and again becomes localized after reflection from the other end of the system. We term this effect `unscrambling', which is marked by the value of the OTOC becoming close to 1
again. Interestingly, this feature 
disappears gradually with increasing interactions. It is not well-understood yet if the presence 
or absence of unscrambling is due of the non-integrability of the system or only because 
of interactions between the excitations.

\begin{figure}[H]
\centering
\includegraphics[width=0.48\textwidth]{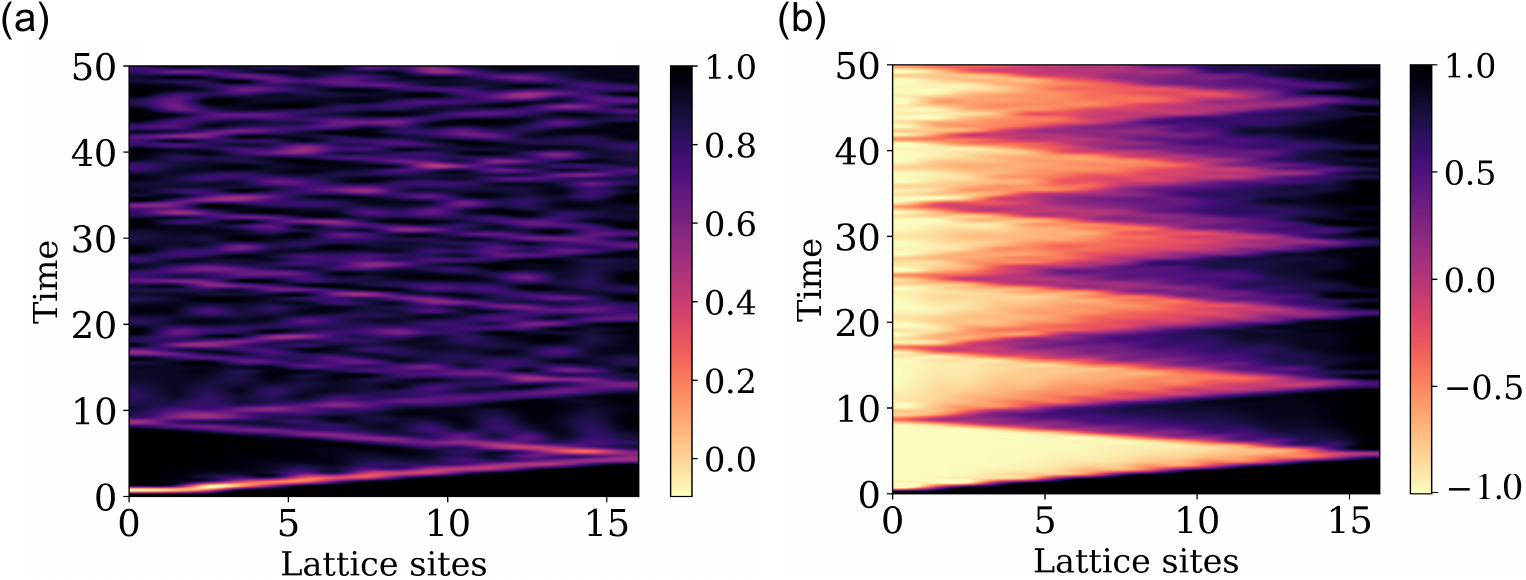}
\caption{Spatiotemporal evolution of the ground state OTOC for (a) the JW
local operator $\si^{z}$ and (b) the JW non-local operator $\si^{x}$, at 
the critical point of the TFIM given in Eq.~\eqref{Ising_ham} with $J_x = h =2$. The $\si^{z}$-OTOC shows only light-cone
lines of information propagation, while the $\si^{x}$-OTOC 
shows alternating scrambling of quantum information and unscrambling after reflections from the ends.} \label{fig:Sz-Sx_compare} \end{figure}

In an attempt to understand the mechanism of scrambling and unscrambling of operators in terms of the nature of excitations of the system, we study the ground state OTOC of JW 
non-local operators (local magnetization) in the transverse field Ising model (TFIM) with and without interactions. The OTOC for JW local operators in the TFIM, which shows only propagation of the 
front without scrambling, can be evaluated exactly using Wick's theorem. But the calculation of the OTOC for JW {\it non-local} operators (see Fig.~\ref{fig:Sz-Sx_compare}) in 
the fermionic language is tricky and prevents a simple physical understanding of the
effects mentioned above. Therefore, we resort to a perturbative approach directly in the 
spin language. We demonstrate that in the paramagnetic (PM) phase of the TFIM, the operator creates spin-flip excitations above the ground state. The scrambling and
unscrambling of quantum 
information happens due to the scattering phase shifts of two spin-flip excitations. Conversely, in the ferromagnetic (FM) phase, the operator excites domain walls~\cite{suchsland2022} above the ground 
state which are responsible for the scrambling and unscrambling. 
Additionally, if interactions are added, the unscrambling in the ground state OTOC
fades away. We find that this absence of unscrambling is due to different 
reasons in the PM and the FM phases. In the PM phase, even a small interaction alters the scattering phase shift significantly from the non-interacting case and obstructs unscrambling. In the FM phase, a comparatively larger interaction is needed to
disrupt the unscrambling by creating higher order domain wall excitations.
In this letter, we will provide a detailed quantitative analysis of this mechanism with 
a comparison of the OTOCs calculated using numerical exact diagonalization and our effective theory of low-energy excitations. The excellent match between the
two confirms that the operator spreading in the ground state is indeed governed 
by the low-energy excitations. 

{\it Ground state OTOC.} The Hamiltonian of the TFIM is 
\begin{equation}
H ~=~ - ~J_x ~\sum_{n=0}^{L-2} \si^{x}_{n} \si^{x}_{n+1} ~-~ h~ \sum_{n=0}^{L-1}\si^{z}_{n}, \label{Ising_ham} \end{equation}
on a lattice with $L$ sites and open boundaries. As the transverse field $h$ is lowered from a large value, the model is known to exhibit a continuous phase transition from paramagnetic to ferromagnetic at $h = J_x$. We consider the OTOC of the local magnetization operator $\si^{x}$ starting from one end, $F^{xx}(l,t) = \langle \si^{x}_{l}(t) \si^{x}_{0} 
\si^{x}_{l}(t) \si^{x}_{0} \rangle $. We will analyze the OTOCs in the PM 
and FM phases using a perturbative treatment for large and small 
transverse fields respectively.

\begin{figure}[H]
\centering
\includegraphics[width=0.48\textwidth]{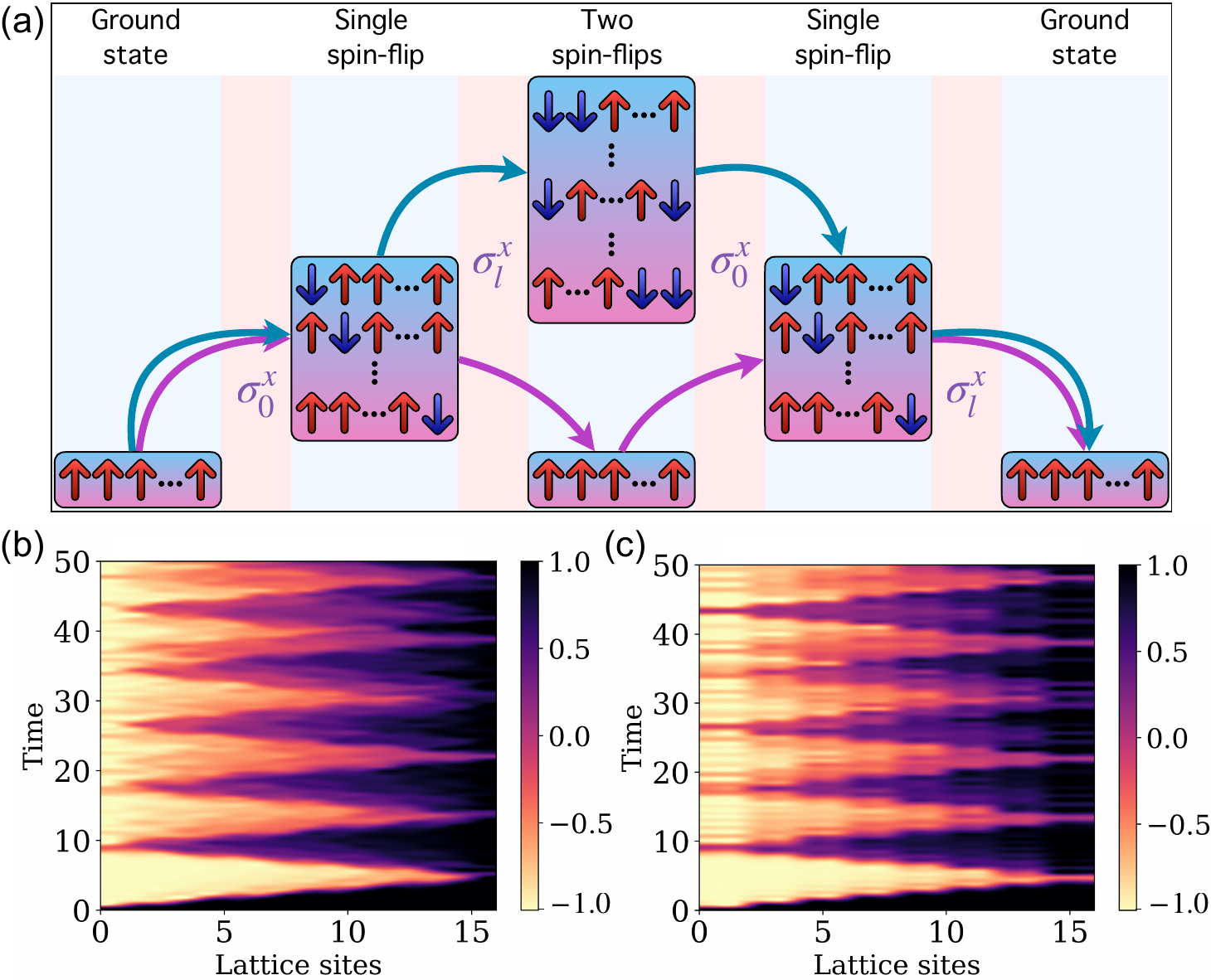}
\caption{(a) Schematic of the mechanism of scrambling and unscrambling of $F^{xx}
(l,t)$ involving two spin-flip excitations in the paramagnetic phase of the TFIM. 
(b) Spatiotemporal evolution of $F^{xx}(l,t)$ as 
obtained from a numerical time evolution for $J_x = 2$ and $h =4$. (c) $F^{xx}(l,t)$ for the same parameter values obtained using the effective theory of two spin-flip excitations as mentioned in the 
text. The exact numerics and the effective theory calculations give remarkably similar results for the scrambling and unscrambling of $\sigma^x$.} \label{fig:spin_flip} \end{figure}

{\it Paramagnetic phase: spin-flip excitations.}
In the paramagnetic phase, at large values of the transverse field, $|h/J_x| \gg 1$, the ground state of the model has all the spins aligned in the direction of the field, $\ket{\uparrow \uparrow \uparrow
\cdots \uparrow}_{z}$ in the $\si^{z}$-basis. The lowest energy excitations are given
by single spin-flips like $\ket{\uparrow \downarrow \uparrow \cdots \uparrow}_{z}$. In
the OTOC, $\si^{x}_{0}$ acting on the ground state creates a spin-flip excitation at the left-most (zero-th) site which can then hop through the chain due 
to the $J_x\si^{x}_{n} \si^{x}_{n+1}$ term in the Hamiltonian. Subsequently, $\si^{x}_{l}$ can either produce a second spin-flip excitation or de-excite the system 
back to the ground state. Thus, the ground state OTOC is approximately governed by a combination of two processes, one which involves only single spin-flip 
excitations and the other involving two spin-flips (see Fig.~\ref{fig:spin_flip} (a)). 
This becomes clear if we write the OTOC using the intermediate excited 
states by resolving the identity approximately as $\mathbb{I} \approx \ket{\psi_{GS}}
\bra{\psi_{GS}} + \sum_{q} \ket{q} \bra{q} + \sum_{q_1,q_2} \ket{q_1, q_2} 
\bra{q_1,q_2} + $ higher order excitations. 
\begin{align}
&F^{xx}(l,t) \simeq \nonumber \\ & \sum_{q, q'} \big[ \bra{\psi_{GS}} e^{iHt} \si^{x}_{l} \ket{q'} \bra{q'} e^{-iHt} \si^{x}_{0} \ket{\psi_{GS}} 
\bra{\psi_{GS}} e^{iHt} \si^{x}_{l} \ket{q} \nonumber \\ 
&\bra{q} e^{-iHt} \si^{x}_{0} \ket{\psi_{GS}}\big] + \sum_{q_1, q_4, q_2 < q_3} \big[\bra{\psi_{GS}} e^{iHt} \si^{x}_{l} \ket{q_4} \nonumber \\ &\bra{q_4} e^{-iHt} 
\si^{x}_{0} \ket{q_2, q_3} \bra{q_2, q_3} e^{iHt} \si^{x}_{l} \ket{q_1} \bra{q_1} e^{-iHt} \si^{x}_{0} \ket{\psi_{GS}}\big], \label{spin_flip_OTOC1}
\end{align}
where $\ket{q}$ and $\ket{q_1,q_2}$ denote the eigenstates of the Hamiltonian with single and two spin-flip excitations, respectively. The quantum number $q$ takes the values $0,1, 
\dots,L-1 $. It is worth noting that it is enough to consider processes up to the second order if $J_x /h$ is small. The matrix elements in Eq.~\eqref{spin_flip_OTOC1} can be evaluated using the 
energies and wave functions of the excited states $\ket{q}$ and $\ket{q_1,q_2}$. 
The single spin-flip excitations are described by an effective nearest-neighbour 
tight-binding model on the $L$-site lattice with open boundaries, and therefore have
energies $\varepsilon_{q} = -2J_x~ \cos(\frac{\pi q}{L+1})+2h$ above the ground
state, and wave functions $\psi_{q}(n) = \braket{n|q} = \sqrt{2/(L+1)}~ \sin[\pi 
(q+1) (n+1)/(L+1)]$, where $\ket{n} = \si^{x}_{n} \ket{\psi_{GS}}$ denotes the state 
with a single flipped spin at site $n$. 

While the single-particle eigenstates are solved by considering the problem of a particle in a box on a lattice, the case of two spin-flips is a bit subtle. The two spin-flips 
should be on separate sites and follow commutation relations, making it a system 
with two hard-core bosons. Due to the indistinguishability, we can write the wave 
function for two spin-flip excitations as $\psi_{q_1,q_2}(n_1,n_2) = \braket{n_1, n_2| q_1,q_2} = \psi_{q_1}(n_1)\psi_{q_2}(n_2)- \psi_{q_2}(n_1)\psi_{q_1}(n_2)$, 
where $\ket{n_1,n_2}$ denotes the state with two spin-flips at sites $n_1$ and 
$n_2$, and we choose $n_1 < n_2$. The energy this state is the sum of 
two single-particle energies, $\varepsilon_{q_1,q_2} =\varepsilon_{q_1}+ \varepsilon_{q_2}$. Finally, we arrive at~\cite{SM}
\begin{align}
&F^{xx}(l,t) \simeq \sum_{q,q'} e^{-i(E_q +E_{q'})t} \psi_{q}(l) \psi_{q}(0) \psi_{q'}(l) \psi_{q'}(0) \nonumber \\ & ~+\sum_{q_1,q_4,q_2<q_3} e^{-i(E_{q_1}+E_{q_4}-
E_{q_2} -E_{q_3})t} \psi_{q_1}(0)\psi_{q_4}(l) \nonumber \\ 
& \sum_{n_1} \psi_{q_1}(n_1) \big[\theta(n_1 - l) \psi_{q_2,q_3}(l,n_1) + \theta(l-n_1) \psi_{q_2,q_3}(n_1,l) \big] \nonumber \\
&\sum_{n_2} \psi_{q_4}(n_2) \psi_{q_2,q_3}(0,n_2). \label{spin_flip_OTOC2}
\end{align}

We evaluate this expression exactly and contrast the spatiotemporal evolution of $F^{xx}(l,t)$ calculated using exact numerical time-evolution with this analysis in Figs.~\ref{fig:spin_flip} (b) and (c). The striking agreement confirms that spin-flip excitations indeed provide the mechanism
responsible for the scrambling of quantum information and the unscrambling after 
a reflection when $h$ is large. Interestingly, this analysis agrees quite well
with the exact result even if $h$ is not much larger than $J_x$. In fact, the plot for OTOC looks 
qualitatively the same for $h= J_x$ (Fig.~\ref{fig:Sz-Sx_compare} (b)) and $h >J_x$ (Fig.~\ref{fig:spin_flip} (b)), although an analysis in terms of the spin-flip excitations cannot be assumed to be valid at the critical point.

\begin{figure}[H]
\centering
\includegraphics[width=0.48\textwidth]{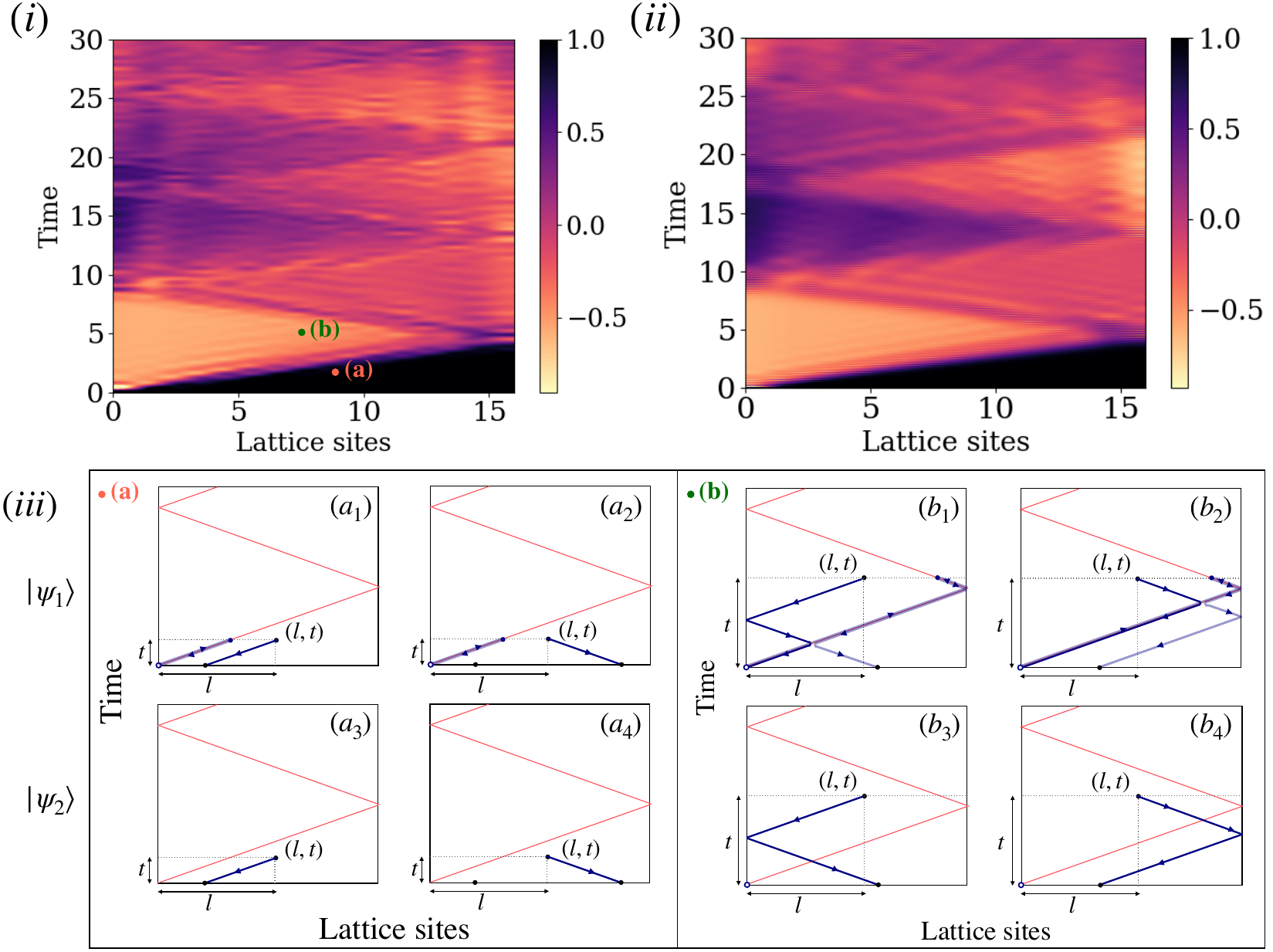}
\caption{Spatiotemporal plots of $F^{xx}(l,t)$ for the interacting TFIM with $J_x = 2, ~h = 8, ~J_z = 1.8$ obtained $(i)$ from exact numerics and $(ii)$ from the analysis using two spin-flip excitations. The value of $J_z$ is such that a region which showed scrambling in the non-interacting case now shows OTOC close to 1. $(iii)$ The point $(a)$ in the space-time plot corresponds to a region where there is no phase difference due to scattering between the states $\ket{\psi_1}$ and $\ket{\psi_2}$ (left panel), while at point $(b)$ (right panel) there is a scattering event between the two excitations in $\ket{\psi_1}$, and therefore the OTOC, defined as the inner product $\braket{\psi_2|\psi_1}$ picks out the phase difference between the states.} \label{fig:Large_h_interaction}
\end{figure}

{\it Interactions in the presence of large field.}
We now look at the behavior of the OTOC in the PM phase in the presence of an integrability-breaking interaction term given by $J_z \sum_{n} \si^{z}_{n} \si^{z}_{n+1}$. The full Hamiltonian is given by
$H = -J_x \sum_{n=0}^{L-2} \si^{x}_{n} \si^{x}_{n+1}- h \sum_{n=0}^{L-1}
\si^{z}_{n} + J_z \sum_{n=0}^{L-2} \si^{z}_{n} \si^{z}_{n+1}$.
When $|h/J_x| > 1$, a small value of $J_z$ can significantly change the spatiotemporal 
plot for $F^{xx}(l,t)$, as shown in Fig.~\ref{fig:Large_h_interaction} (a). 
As usual, the local quantum information starts scrambling after $\si^x_{0}$ acts at one end. However, we do not see any unscrambling after reflection, unlike in the
model without interactions. As time progresses further, the local information always stays scrambled and never becomes localized to a few sites. 
This can also be explained using a simple picture of two-spin flips. We need to modify Eq.~\eqref{spin_flip_OTOC2} to account for two-spin flip states in the presence of interactions and evaluate it numerically. The wave functions for the two-spin flip state $\psi_{q_1,q_2}(n_1, n_2)$ are replaced by two-body eigenstates calculated numerically by considering an effective tight-binding model with two hard-core bosons with a density-density interaction given by $J_z$. The OTOC calculated using this analysis involving only up to two-spin flips shows fairly good agreement with the plots obtained using exact numerics even with interactions. We contrast the two plots in Figs. ~\ref{fig:Large_h_interaction} $(i)$ and $(ii)$ for the parameter values $J_x = 2, ~h = 8$, and $J_z = 1.8$.

We now introduce a simple space-time picture to understand the various spatiotemporal plots of the OTOC. We rewrite the expression for the OTOC as $F^{xx}(l,t) = \bra{\psi_{GS}} e^{iHt}\sigma^{x}_{l} e^{-iHt} \sigma^{x}_{0}e^{iHt}\sigma^{x}_{l} e^{-iHt} \sigma^{x}_{0} \ket{\psi_{GS}} = \braket{\psi_2| \psi_1}$, where $\ket{\psi_1}$ and $\ket{\psi_2}$ are both states with one particle (spin-flip) and are defined respectively as $\ket{\psi_1} = \sigma^{x}_{0}e^{iHt}\sigma^{x}_{l} e^{-iHt} \sigma^{x}_0 \ket{\psi_{GS}}$ and $\ket{\psi_2} = e^{iHt}\sigma^{x}_{l} e^{-iHt}\ket{\psi_{GS}}$. The state $\ket{\psi_1}$ can be interpreted as follows. Starting from the right, the first $\sigma^{x}_0$ operator creates a spin-flip at the end site which then propagates with some velocity for a time $t$ (due to $e^{-iHt}$). Next another spin-flip is created at site $l$. The two flipped spins then propagate back in time (due to $e^{iHt}$) to $t=0$ which involves zero, one or more scatterings between them depending on the position $l$ and time $t$. Finally, at $t=0$ one of the excitations is annihilated at site $l=0$ to produce a state with only one spin-flip.
In contrast, $\ket{\psi_2}$ has a simpler interpretation as it involves a single spin-flip which is created at position $l$ and time $t$ on the time-evolved ground state, and which then propagates back in time to $t=0$. The inner product $\braket{\psi_2|\psi_1}$ then depends on the scattering phase shift in the state $\ket{\psi_1}$ with respect to state $\ket{\psi_2}$. 

Some assumptions are now required to estimate the value of the OTOC using scattering phase shifts. First, even though we are considering a system with open boundaries, we will assume that the system is large enough such that away from the edge, the momentum $q$ is a good quantum number. Second, we approximate a spin-flip excitation to be a quasiparticle with a single momenta $q_0$ with the largest velocity given by the Lieb-Robinson bound, although actually it is a superposition of many momentum states around $q_0$, propagating with the maximum possible group velocity. The entire region in the spatiotemporal plot of the OTOC can now be divided into regions according to the number of scatterings between the pair of quasiparticles with momenta $q_0$ and $-q_0$ in the state $\ket{\psi_1}$. The value of the OTOC (real part) in each region is approximately constant and is given by the real part of the scattering phase shift obtained from the two-particle Bethe ansatz \cite{SM}.

The scrambling and unscrambling for the TFIM without interactions, given by the successive dark (OTOC close to $1$) and bright (OTOC close to $-1$), regions can be now understood as the regions having even and odd number of scatterings respectively starting from the lowest dark region with zero scattering. In the absence of interactions, every scattering event changes the phase by $e^{i\theta} = -1$. For the interacting model, the absence of unscrambling can be simply understood as the deviation of scattering phase shifts from $-1$ due to $J_z$. However, it should be noted that for later times the contributions from excitations not corresponding to the largest group velocity become significant and they alter the value of the OTOC obtained from the estimate given by scattering phase shift of two excitations.

To illustrate this, we have chosen two points $(a)$ and $(b)$ in the OTOC plots in Fig.~\ref{fig:Large_h_interaction} $(i)$ where point $(a)$ is in a region of no scrambling and $(b)$ is in a region of scrambling before any reflection. We now consider the schematic shown in Fig.~\ref{fig:Large_h_interaction} $(iii)$. The red lines are guides to the eye for the light-cone front. Then $\ket{\psi_1}$ at $(a)$ is given by a superposition of one-particle states shown in $(a_1)$ and $(a_2)$ with an intermediate process where a second excitation is first created at $(l,t)$ = $(0,0)$, then propagated forward in time, propagated back in time, and finally annihilated at the same point of creation. By contrast, $\ket{\psi_2}$ at $(a)$ is given by superposition of one-particle states shown in $(a_3)$ and $(a_4)$ without any other particles created or annihilated. Since there is no scattering event for point $(a)$, $F^{xx} = \braket{\psi_2|\psi_1} =1$. Similarly, for point $(b)$, $\ket{\psi_1}$ is made of a superposition of states shown in $(b_1)$ and $(b_2)$ and $\ket{\psi_2}$ is made of a superposition of states shown in $(b_3)$ and $(b_4)$. In contrast to point $(a)$, we have one scattering phase shift between two excitations in the state $\ket{\psi_1}$ with respect to the state $\ket{\psi_2}$. 
Hence the inner product of the two states is given by the scattering phase shift
which, for a pair of non-interacting hard-core bosons, is equal to $-1$; this explains the value of $F^{xx}(l,t)$ in the first scrambled region in the non-interacting TFIM. For the interacting case also it has a value close to $-1$. Subsequently, for larger times the scrambled and unscrambled regions have values of $F^{xx}(l,t)$ given by the total number of scattering events multiplied by the scattering phase shifts of one scattering event. In Figs.~\ref{fig:Large_h_interaction} $(i)$ and $(ii)$ we have considered a $J_z$ so that the region corresponding to three scattering events has $F^{xx}(l,t)$ close to $1$, which, for the non-interacting case, would have been close to $-1$. 

The same effects are observed if we consider the integrable but interacting $XXZ$ spin-1/2 model in a transverse field. Namely, for the Hamiltonian, $H' = -J \sum_{n=0}^{L-2} 
\si^{x}_{n} \si^{x}_{n+1} -J \sum_{n=0}^{L-2} \si^{y}_{n} \si^{y}_{n+1} - 
h \sum_{n=0}^{L-1}\si^{z}_{n} + J_z \sum_{n=0}^{L-2} \si^{z}_{n} 
\si^{z}_{n+1}$, 
the spatiotemporal plot for the OTOC shows a similar behavior including the
absence of unscrambling as compared to the non-interacting $XX$ model without the $J_z$-term (see Fig.~\ref{fig:supplement_XXZ} in Ref.~\cite{SM}).
Therefore, a similar analysis in terms of the low-energy excitations holds true for the $XXZ$ spin-1/2 model also, implying that the absence of unscrambling is not related to the the integrability or non-integrability of the model.

\begin{figure}[h]
\centering
\includegraphics[width=0.48\textwidth]{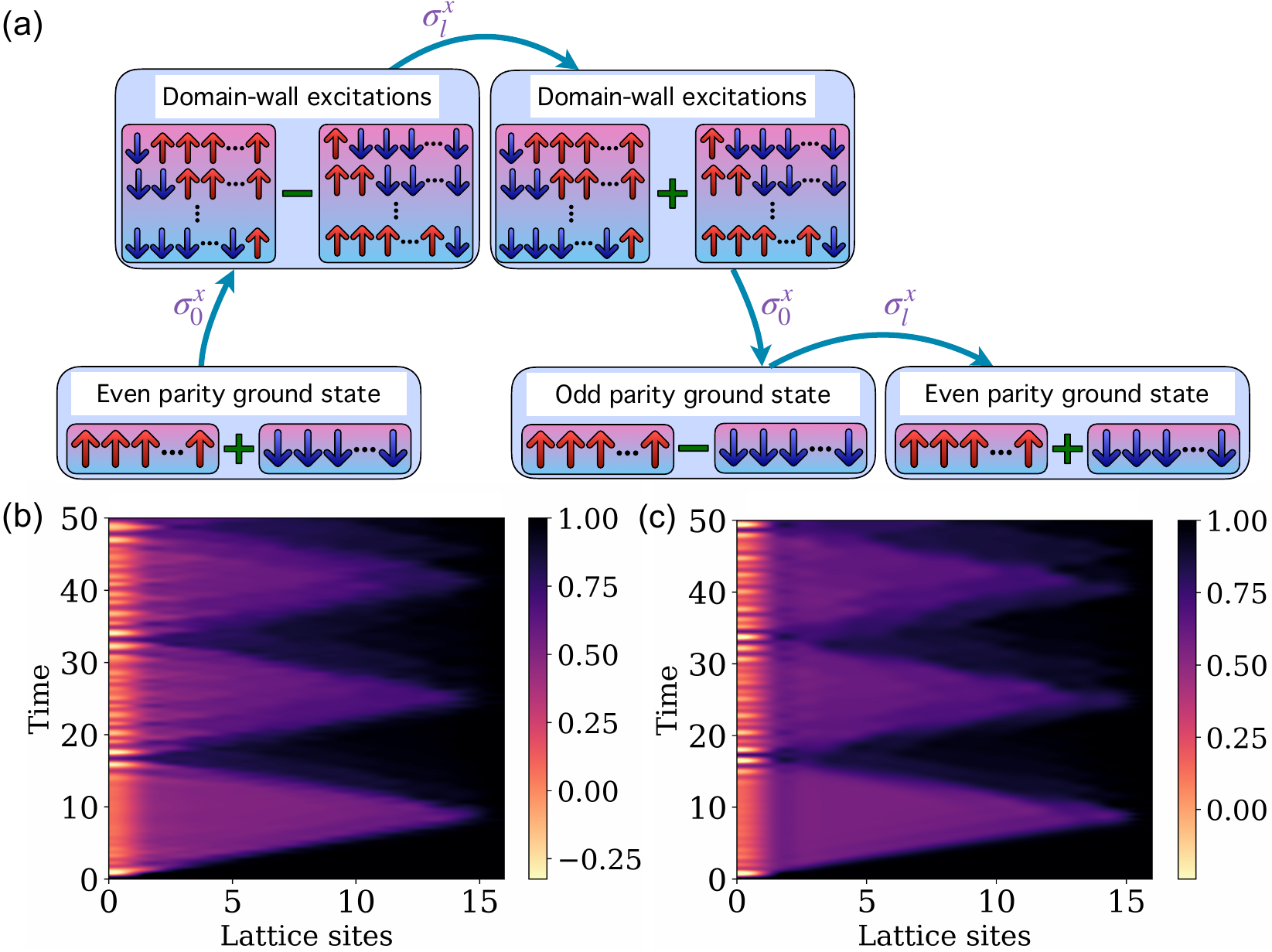}
\caption{(a) Schematic of the mechanism of domain-wall assisted scrambling and 
unscrambling in the ordered phase of the TFIM. Spatiotemporal plots of 
$F^{xx}(l,t)$ for $J_x =2$ and $h=1$ obtained (b) using exact numerics in the ordered phase, and (c) using the analysis of 
domain-wall excitations described in the text. The remarkable agreement between these two plots suggests that indeed scrambling and unscrambling in the FM phase is governed by domain-wall excitations.}
\label{fig:domain_wall}
\end{figure}

{\it Ferromagnetic phase: domain walls.}
We now discuss the scrambling in the FM phase of the TFIM. For very small $h$, the ground states are the degenerate states $\ket{I} = \ket{\uparrow \uparrow \cdots \uparrow}_{x}$ and $\ket{II} = \ket{\downarrow \downarrow 
\cdots \downarrow}_{x}$ in the $\si^{x}$-basis. The model has a $Z_2$ spin-flip symmetry, but the ground states spontaneously break it and the true ground states are the superpositions 
$\ket{+} = (\ket{I}+\ket{II})/\sqrt{2}$ and $\ket{-} = (\ket{I}-\ket{II})/\sqrt{2}$ corresponding to the even and odd fermion parity respectively. Setting $J_x = 1$, 
and taking $h$ small, a perturbative expansion up to first order in $h$~\cite{SM} yields the modified ground states expressed in a product form $\ket{I'} = \ket{(\uparrow+\frac{h}{2}\downarrow)_{0} (\uparrow+\frac{h}{4} \downarrow)_{1} \cdots (\uparrow+\frac{h}{2}\downarrow)_{L-1}}_{x}$ and $\ket{II'} = \ket{(\downarrow+\frac{h}{2}\uparrow)_{0} (\downarrow+\frac{h}{4}\uparrow)_{1} \cdots (\downarrow+\frac{h}{2}\uparrow)_{L-1}}_{x}$ (up to a normalization constant), where the subscripts denote the site indices. The lowest energy excitations are the two types of domain walls, namely $\ket{m+1/2,I}$ and $\ket{m+1/2,II}$, where the site label $m+\frac{1}{2}$ implies that the domain wall is situated between the $m$-th and $(m+1)$-th sites, and $I$ ($II$) denotes that all the spins on the left of the domain wall are down (up). Here $m$ can take values from $0$ to $L-2$. 
For example, $\ket{0+1/2, I} = \ket{\downarrow \uparrow \uparrow \cdots \uparrow}_{x}$ denotes the state with a domain wall between the first and second spins. Now, $\si^{x}_{0}$ acting
on the state $\ket{I'}$ creates the domain-wall state $\ket{0+1/2, I}$ with an amplitude $h$ to leading order. Under time evolution, this domain wall can
move through the lattice 
with one less site due to the $h\sum \si^{z}_{n}$ term in the Hamiltonian, allowing us to construct 
domain-wall eigenstates with a quantum number label 
$\kappa$. Accordingly, domain-wall states have an energy dispersion $\epsilon_{\kappa, \alpha} = -2h$ sin$[\pi (\kappa+1)/L] +2J_x $ above the ground state, and a wave 
function $\phi_{\kappa,\alpha}(m) = \braket{m+1/2,\alpha \mid \kappa, \alpha} 
= \sqrt{2/L}~$sin$[\pi (\kappa+1) (m+1)/L]$, where $\kappa = 0,1, \cdots, L-2$, and $\alpha$ can be $I$ or $II$. Subsequently, $\si^{x}_{l}$ for $l \neq 0$
does not create or destroy the domain-wall excitation but only measures if the spin 
is up or down in the $x$-basis. However, it changes the parity of the domain-wall excitation. 
The remaining $\si^{x}_{0}$ and $\si^{x}_{l}$ operators first de-excite the domain 
wall to a ground state with the opposite parity and then returns it to the ground state
we started with. In short, as the domain moves from one end to the other end, it leaves flipped spins behind along its trajectory. This is then measured by the operator $\sigma^{x}_{l}$ at different times. The entire mechanism is described schematically in Fig.~\ref{fig:domain_wall} (a). We have a different situation
when $l=0$, as the four $\si^{x}_{0}$ operators sequentially excite and de-excite between the ground state and a domain-wall state, resulting in a different value of the OTOC at the edge. 

We now present a semi-analytical treatment of domain-wall dynamics to quantitatively understand scrambling and unscrambling. To this end, we construct the $\si^{x}$ operators 
at all sites which is perturbatively correct up to order $h$
in the subspace consisting only of the ground state and single 
domain-wall states. The operator $\si^{x}_{0}$ has non-trivial matrix elements only in the subspace of states $\{\ket{I'}, \ket{0+1/2, 
I}\}$ and $\{\ket{II'}, \ket{0+1/2, II}\}$. On the other hand, $\si^{x}_{L-1}$ has non-trivial elements in the subspaces spanned by the states $\{\ket{I'}, \ket{L-2+1/2, II}\}$ 
and $\{\ket{II'}, \ket{L-2+1/2, I}\}$.
The $\si^{x}_{l}$ for all other sites are diagonal and have trivial matrix elements ($\pm 1$) up to order $h$ as they only determine whether the spin is up or down. The 
Hamiltonian then has a simple structure with only the ground state energy along the diagonal in the ground state basis and a tight-binding structure in the subspace of domain 
walls. Since the Hamiltonian only appears in the exponential in time evolution, we do not need to consider the matrix elements between the ground states and the domain walls to the lowest order. 

The effective representations of $\si^{x}_{n}$ and the Hamiltonian $H$ in the 
subspace spanned by the set of states $\{\ket{I'}, \{\ket{m+1/2,I}
\}_{m=0}^{L-2}, \ket{II'}, \{\ket{m+1/2, II}\}_{m=0}^{L-2}\}$ are defined as $X_{n}$ 
and $\mathcal{H}$ (see Ref.~\cite{SM} for the explicit matrix forms). The OTOC in 
terms of domain-wall excitations can be written as
\begin{align}
&F^{xx}(l,t) \approx \nonumber \\
&\frac{1}{2}\bra{I'+II'} e^{i\mathcal{H}t} X_{l}e^{-i\mathcal{H}t} X_{0} e^{i\mathcal{H}t} X_{l}e^{-i\mathcal{H}t} X_{0} \ket{I'+II'}. \label{OTOCh<1}
\end{align}
The plots for spatiotemporal evolution of $F^{xx}(l,t)$ using exact numerics and our analysis agree very well as shown in Figs.~\ref{fig:domain_wall} (b) and (c). 
Moreover, in both the plots we observe the reflection of scrambled quantum information happening slightly before it reaches the other end, which validates 
that it is indeed governed by domain-wall excitations (which are defined midway between sites). The OTOC expression given in Eq.~\eqref{OTOCh<1} is correct up to order $h^2$.

\begin{figure}[H]
\centering
\includegraphics[width=0.48\textwidth]{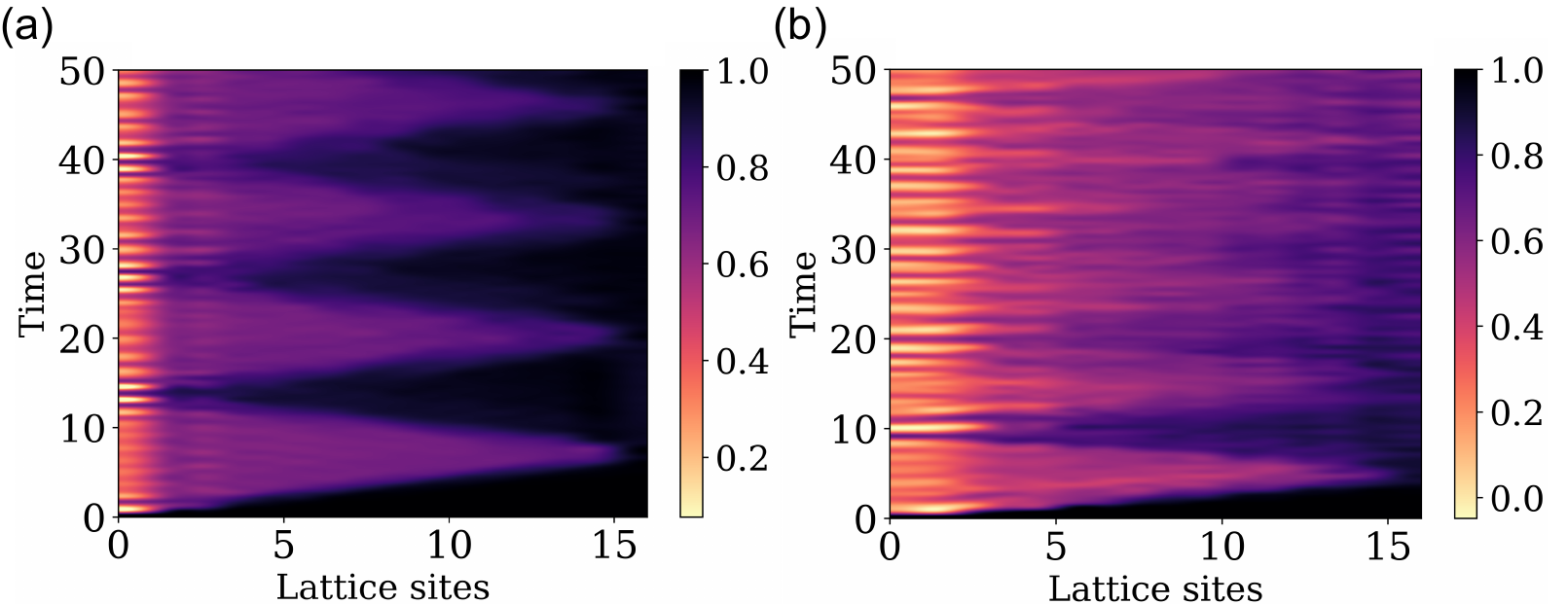}
\caption{$F^{xx} (l,t)$ showing a pronounced unscrambling effect for (a) $J_x =2, ~h=1, ~J_z = 0.5$, while for (b) $J_x =2, ~h=1, ~J_z = 1$, it shows much less unscrambling. As $J_z$ becomes comparable to $J_x$ the perturbative expansion around the ground state becomes invalid. Also, higher order excitations become important and the scrambling is no longer due to single domain-wall motion.}
\label{fig:small_h_interaction} \end{figure}

{\it Interactions in the presence of small field.}
In the FM phase, the introduction of a small interaction $J_z$ (relative to $J_x$) in the Hamiltonian does not alter the qualitative features of the $F^{xx}(l,t)$ much. We can see scrambling and unscrambling much like the case without interactions in the spatiotemporal plots in OTOC. A small $J_z$ also gives rise to dynamics of a domain-wall by making it hop by two sites, in contrast to the transverse field $h$ which leads to hopping by one site. Therefore, the presence of a small interaction effectively only changes the dispersion of the domain walls to $E_k = - 2 h \cos k + 2 J_z \cos (2k)$. The group velocity is then given by $v_k = dE_k /dk = 2 h \sin k - 4 J_z \sin (2k)$. As the light-cone velocity is given by the maximum group velocity of the quasiparticles, for small $J_z/J_x$, the interacting model (Fig.~\ref{fig:small_h_interaction} (a)) shows a different light-cone velocity than its non-interacting counterpart. 

If $J_z$ is increased further, we find that the unscrambling starts to go away. Roughly, this happens around $J_z/J_x \gtrsim 0.4 $, as observed numerically. Fig. ~\ref{fig:small_h_interaction} (b) shows the absence of unscrambling for the parameter values $J_x =2, ~h=1, ~J_z = 1$.
It can be reasonably expected that for such  values of $J_z$, the perturbative expansion starting from the states $\ket{I}$ and $\ket{II}$ will no longer be accurate. Furthermore, the assumption that only the lowest order excitations are responsible for information scrambling also becomes a gross oversimplification. To explain the absence of unscrambling when the interaction $J_z$ is comparable with $J_x$, one needs to account for higher order excitations like dynamics of three or more domain walls.

{\it Conclusion.}
To summarize, we have studied the propagation of local quantum information in a spin-1/2 chain with open boundary conditions using ground state OTOCs
of $\si^z$ and $\si^x$ operators which are local and non-local in terms of JW fermions respectively. While both $\si^z$ and $\si^x$ OTOCs show light-cone
like propagation, only the $\si^x$ OTOC 
shows scrambling within the light-cone, in agreement with earlier results. In addition, we discover remarkable unscramblings and scramblings of the $\si^x$ OTOC after repeated reflections from the ends. We have provided, for the first time,
an analytical understanding of both scrambling and unscrambling deep
in the PM phase (when the transverse field $h$ is large) and the FM phase 
(when the $XX$ coupling $J_x$ is large) phases in terms of the low-energy excitations, namely, spin-flips when $h$ is large and domain walls when $J_x$
is large.
When an interaction between JW fermions ($ZZ$ coupling) is added, the 
unscrambling effect becomes weaker. In the PM phase, even relatively
weak interactions significantly change the scattering phase shift for 
two spin-flips and thereby reduces the unscrambling. In the FM phase, 
stronger interactions are required to destroy unscrambling and it occurs
due the the creation of multiple domain wall excitations.

To conclude, we point out a recent experimental measurement of the OTOC \cite{green2022} at finite temperature which, as the authors suggest, can also be performed for the ground state. In this paper, finite temperature OTOCs of the transverse field Ising model are studied for a trapped linear chain of $^{171}$Yb$^{+}$ ions by creating a thermofield double state and then looking at
its time evolution. We believe that a similar route can be 
followed to experimentally study the dynamics of low-energy excitations through 
the ground state OTOCs of the spin chains discussed in our work.

{\it Acknowledgements.}
S.S. thanks MHRD, India for financial support through the PMRF.
D.S. acknowledges funding from SERB, India (JBR/2020/000043).

\bibliography{references}

\newpage
\begin{widetext}

\section{Supplemental Material}

\section*{A: Derivation of expression for OTOC in Eq.~\eqref{spin_flip_OTOC2}}

We will present here the derivation of the final expression of the OTOC in terms of 
spin-flip excitations given in Eq.~\eqref{spin_flip_OTOC2} for $|h/J_x| \gg 1$.
The time evolution of the spin-flip eigenstates $e^{-iHt} \ket{q}$ and $e^{-iHt} \ket{q_1,q_2}$ is trivial. The more difficult part is to find the matrix elements 
$\bra{q_1,q_2} \si^{x}_{l} \ket{q}$. We use the resolution of identity,
\begin{align}
\bra{q_2,q_3} \si^{x}_{l} \ket{q_1} =\big(\frac{2}{L+1} \big)^{3/2} \sum_{n_1, n_2<n3} 
\braket{q_2,q_3|n_2,n_3} \braket{n_2, n_3| \si^{x}_{l} | n_1} \braket{n_1|q_1},
\tag{S1}
\end{align}
and the wave functions $\braket{q_2,q_3|n_2,n_3} = \psi_{q_1,q_3}(n_2,n_3)$ and $\braket{n_1|q_1} =\psi_{q}(n)$ as given in the main text. The matrix element 
$\braket{n_2, n_3| \si^{x}_{l} | n_1}$ depends on whether $l < n_1$ or $l>n_1$ and we always choose $n_2<n_3$. Since $\si^{x}_{l}$ creates a spin-flip, we should have $l \neq 
n_1$. Therefore,
\begin{align*}
\braket{n_2, n_3| \si^{x}_{l} | n_1} = \theta(n_1-l) \delta_{n_2,l} \delta_{n_1,n_3} 
+ \theta(l-n_1) \delta_{n_3,l} \delta_{n_1,n_2}. 
\tag{S2}
\end{align*}
Using these relations we obtain Eq.~\eqref{spin_flip_OTOC2} in the main text.

\section*{B: Perturbative ground state for $|h/J_x| <1$}
k
The degenerate ground states of the model in the limit of $h \to 0$ are $\ket{I} = \ket{\uparrow \uparrow \cdots \uparrow}_{x}$ and $\ket{II} = \ket{\downarrow \downarrow 
\cdots \downarrow}_{x}$. If a small $h$ is turned on, we obtain, from degenerate perturbation theory,
\begin{align}
\ket{I'} & = \ket{I} + h\sum_{n} \frac{\ket{n}\braket{n|\sum_{l} \si^{z}_{l}| I}}{ 2(\delta_{n,0} +\delta_{n,N-1}) +4 (1- \delta_{n,0} - \delta_{n,N-1})} ~+~ O(h^2) \nonumber \\ 
& = \ket{I} + \frac{h}{2} \ket{\downarrow \uparrow \uparrow \cdots}_{x} + \frac{h}{4} \ket{\uparrow \downarrow \uparrow \cdots}_{x} + \frac{h}{4} \ket{ \uparrow \uparrow 
\downarrow \uparrow \cdots}_{x} + \cdots + \frac{h}{2} \ket{\uparrow \uparrow 
\cdots \downarrow}_{x},
\tag{S3}
\end{align}
where $\ket{n}$ denotes the single spin-flip state
in the $\si^{x}$-basis at site $n$.
Combining these together we can write down the following perturbative ground state which is valid up to $O(h)$,
\begin{align}
\ket{I'} = \ket{(\uparrow+\frac{h}{2}\downarrow)_{0} (\uparrow+\frac{h}{4}\downarrow)_{1} \cdots (\uparrow+\frac{h}{2}\downarrow)_{L-1}}_{x}.
\tag{S4} 
\end{align}
In a similar way, we can derive the other perturbed ground state,
\begin{align}
\ket{II'} = \ket{(\downarrow+\frac{h}{2}\uparrow)_{0} (\downarrow+\frac{h}{4}\uparrow)_{1} \cdots (\downarrow+\frac{h}{2}\uparrow)_{L-1}}_{x}.
\tag{S5}
\end{align}

\section*{C: Matrix forms for the operators in the domain-wall picture}

We will derive here the matrix forms of the $\si^{x}$ 
operators in the ground state and the single domain-wall sector.
We start with the $\si^{x}_{0}$ operator acting on the perturbed ground state $\ket{I'}$, we will 
always make sure to consider processes which lie in low-energy sector and to work
only up to $O(h)$.
\begin{align}
\si^{x}_{0} \ket{I'} = \ket{(\uparrow-\frac{h}{2}\downarrow)_{0}(\uparrow+\frac{h}{4}\downarrow)_{1} \cdots (\uparrow+\frac{h}{2}\downarrow)_{L-1}}_{x} = \ket{I'} - 
h\ket{0+1/2,I}.
\tag{S6}
\end{align}
We now consider the perturbed domain-wall state $\ket{0+1/2,I}$ up to an $O(h)$ correction. This can be written compactly as 
\begin{align}
\ket{0+1/2,I'}=\ket{(\downarrow-\frac{h}{2}\uparrow)_{0}(\uparrow)_{1} (\uparrow+\frac{h}{4}\downarrow)_{2} \cdots (\uparrow+\frac{h}{2}\downarrow)_{L-1}}_{x}.
\tag{S7}
\end{align}
Then the action of $\si^{x}_{0}$ on this state yields
\begin{align}
\si^{x}_{0} &\ket{0+1/2,I'} = -\ket{(\downarrow)_{0}(\uparrow)_{1} (\uparrow+\frac{h}{4}\downarrow)_{2} \cdots (\uparrow+\frac{h}{2}\downarrow)_{L-1}}_{x} +h 
\ket{(\uparrow)_{0}(\uparrow+\frac{h}{4}\downarrow)_{1} \cdots (\uparrow+\frac{h}{2}\downarrow)_{L-1}}_{x}.
\tag{S8} 
\end{align}
Up to order $h$ we can simply write
\begin{align}
\si^{x}_{0} \ket{0+1/2,I'} = -\ket{0+1/2,I'} +h\ket{I'}.
\tag{S9} 
\end{align}
Therefore, in this small basis $\{\ket{I'}, \ket{0+1/2,I'} \}$ the matrix form of $\si^{x}_{0}$ is given by
\begin{align}
(X_{0})_{I} \approx
\begin{pmatrix} 
\frac{1}{\sqrt{1+h^2}} & -\frac{h}{\sqrt{1+h^2}} \\
-\frac{h}{\sqrt{1+h^2}} & -\frac{1}{\sqrt{1+h^2}} \\
\end{pmatrix},
\tag{S9}
\end{align}
where we have normalized the matrix to have $X_{0}^{2} =\mathbb{I}$ as it is an approximate 
representation of the unitary operator $\si^{x}_{0}$. Similarly, it can be shown that 
the effective normalized matrix form for the $\si^{x}_{0}$ operator in the basis $\{\ket{II'}, \ket{0+1/2,II'} \}$ is given by,
\begin{align}
(X_{0})_{II} \approx
\begin{pmatrix} 
-\frac{1}{\sqrt{1+h^2}} & \frac{h}{\sqrt{1+h^2}} \\
\frac{h}{\sqrt{1+h^2}} & \frac{1}{\sqrt{1+h^2}} \\
\end{pmatrix}.
\tag{S10}
\end{align}

On the rest of the domain-wall states, $\si^{x}_{0}$ does not produce any low-energy 
excitations up to $O(h)$. Therefore
$\si^{x}_{0}$ only has diagonal matrix elements 
which can be expressed as 
\begin{align}
\braket{m+1/2,\alpha|X_{0}| m+1/2, \alpha} = (-1)^{\alpha}, \text{for} ~m> 0.
\tag{S11}
\end{align}
Here, $\alpha$ is equal to $I, II$ when it appears as a label in the domain-wall state and is equal to $1,2$ when it occurs in the power.

Combining all these, the $X_{0}$ matrix can be written as
\begin{align}
X_{0} &= \frac{1}{\sqrt{1+h^{2}}} \big(\ketbra{I'}{I'} -h \ketbra{I'}{0+1/2,I} 
-h \ketbra{0+1/2,I}{I'} -\ketbra{0+1/2,I}{0+1/2,I} \big) \nonumber \\
& ~~~~- \frac{1}{\sqrt{1+h^{2}}} \big(\ketbra{II'}{II'} -h \ketbra{II'}{0+1/2,II} 
-h \ketbra{0+1/2,II}{II'} -\ketbra{0+1/2,II}{0+1/2,II}\big) \nonumber \\
& ~~~~- \sum_{m = 1}^{L-2} \ketbra{m+1/2,I}{m+1/2,I} ~+~ \sum_{m = 1}^{L-2} 
\ketbra{m+1/2,II}{m+1/2,II}.
\tag{S12}
\end{align}

The matrix form of the $\si^{x}_{L-1}$ operator at the other end can be derived in a similar fashion. The key thing is to note that $\si^{x}_{L-1}$ connects the 
pair of states $\ket{I'}$ and $\ket{L-2+1/2, II}$ and also the pair $\{\ket{II'}, \ket{L-2+1/2, I}\}$. For the other domain-wall states, the matrix element 
of the operator is diagonal up to $O(h)$,
\begin{align}
\braket{m+1/2,\alpha|X_{L-1}| m+1/2, \alpha} = (-1)^{\alpha+1}~~~ \text{for} ~~~ 0< m < L-1.
\tag{S13}
\end{align}
Therefore the matrix representation of $\si^{x}_{L-1}$ in this low-energy sector is given by
\begin{align}
X_{L-1} &= \frac{1}{\sqrt{1+h^{2}}} \big(\ketbra{I'}{I'} -h \ketbra{I'}{L-2+1/2,II} - h \ketbra{L-2+1/2,II}{I'} -\ketbra{L-2+1/2,II}{L-2+1/2,II} \big) \nonumber \\
& ~~~~-\frac{1}{\sqrt{1+h^{2}}} \big(\ketbra{II'}{II'} -h \ketbra{II'}{L-2+1/2,I} - h \ketbra{L-2+1/2,I}{II'} -\ketbra{L-2+1/2,I}{L-2+1/2,I}\big) \nonumber \\
& ~~~~+\sum_{m = 1}^{L-3} \ketbra{m+1/2,I}{m+1/2,I}
~-~ \sum_{m = 1}^{L-3} \ketbra{m+1/2,II}{m+1/2,II}.
\tag{S14}
\end{align}
The spin operators acting at sites other than the two ends do not produce or destroy a single domain 
wall. They can connect the ground state to states 
with two domain walls, but this
is a higher order process and is therefore not considered in our analysis. Therefore, considering 
only the lowest energy excitations, the matrix elements of $\si^{x}_{l}$ for $l=1,\cdots, 
L-2$ are all diagonal. Hence they can be compactly represented as
\begin{align}
X_{l} = 
&\ketbra{I'}{I'} -\ketbra{II'}{II'} ~+~ \sum_{m=0}^{L-2}(\theta(l-m) +\theta(n-l+1))\ketbra{m+1/2,I}{m+1/2,I} \nonumber \\
& ~~ -~ \sum_{m=0}^{L-2}(\theta(l-m) +\theta(n-l+1))\ketbra{m+1/2,II}{m+1/2,II}.
\tag{S15}
\end{align}

\section*{D: Two-body scattering phase shift from the Bethe ansatz}

We consider the Hamiltonian
\beq H ~=~ \sum_{n=-\infty}^{\infty} ~[- ~J_x ~\si^{x}_{n} \si^{x}_{n+1} ~-~ h~ \si^{z}_{n} ~+~ J_z ~\si^{z}_{n} \si^{z}_{n+1}], \tag{S16} \eeq
where we assume for simplicity that the system is infinitely long.
For $h \gg J_x, ~J_z$, the ground state has all spins pointing up in the 
$\si^z$ basis. Let us call this the vacuum state $| {\rm vac} \ra$. The lowest
excited states are given by single spin-flip states where the spin at site $n$ 
points down while all the other spins point up. We will denote this
state as $| n \ra$, and it is related to the vacuum state as $| n \ra =
c_n^\dag | {\rm vac} \ra$. Here $c_n$ and $c_n^\dag$ are annihilation and
creation operators for hard-core bosons, since $(c_n^\dag )^2 | {\rm vac} \ra = 0$. 
We note that $\si_n^z = 1 - 2 c_n^\dag c_n$.
In terms of these operators, the Hamiltonian can be written as
\beq H_c ~=~ \sum_n ~[2 h ~c_n^\dag c_n ~-~ J_x ~(c_n^\dag c_{n+1} ~+~ 
c_{n+1}^\dag c_n) ~+~ 4 J_z ~c_n^\dag c_n c_{n+1}^\dag c_{n+1}^\dag], 
\tag{S17} \eeq
where we have ignored a constant equal to the ground state energy. We see that
there is an interaction between spin-flip states on neighboring sites whose
strength is given by $4J_z$.

The first excited states consist of single spin-flips. Considering a momentum
eigenstate 
\beq | k \ra ~=~ \sum_n ~e^{i kn} ~| n \ra, \tag{S18} \eeq
we find from Eq.~(S16) that the energy-momentum dispersion is
$E_k = 2h - 2 J_x \cos k$, where $k$ takes values in the range $[-\pi,\pi]$.

We now consider states consisting of two spin-flips. Defining such states
as $| n_1, n_2 \ra = c_{n_1}^\dag c_{n_2}^\dag | {\rm vac} \ra$, where $n_1 < n_2$,
we consider a state with a wave function given by the Bethe ansatz
\beq |k_1, k_2 \ra ~=~ \sum_{n_1 < n_2} ~e^{i (k_1 n_1 + k_2 n_2)} + e^{i (\theta +
k_1 n_2 + k_2 n_1)} ~| n_1 , n_2 \ra. \tag{S19} \eeq
The momentum of this state is $k_1 + k_2$ (since the wave function of the
state $| n_1 + 1, n_2 + 1 \ra$ is $e^{i (k_1 + k_2)}$ times the wave function
of $| n_1, n_2 \ra$), and the energy is 
\beq E_{k_1,k_2} ~=~ 4 h ~-~ 2 J_x ~(\cos k_1 ~+~ \cos k_2). \tag{S20} \eeq
The phase shift $\theta$ appearing in Eq.~(S19) can be found in terms of 
$k_1, ~k_2$ by demanding that $H_c | k_1, k_2 \ra = E_{k_1,k_2} | k_1, k_2 \ra$,
and equating the coefficients of each real space state $| n_1, n_2 \ra$ on the 
two sides of this equation. We discover that
\beq e^{i \theta} ~=~ - ~\frac{\cos ( \frac{k_1 + k_2}{2}) ~+~ \frac{2 J_z}{J_x}
~e^{i (k_2 - k_1)/2}}{\cos ( \frac{k_1 + k_2}{2}) ~+~ \frac{2 J_z}{J_x}
~e^{i (k_1 - k_2)/2}}. \tag{S21} \eeq
Note that in the absence of interactions, $J_z = 0$, we get $e^{i \theta} = -1$
for all values of $k_1, ~k_2$.

In the main text, we are interested in the case where the two momenta $k_1, ~k_2$
are equal and opposite to each other. We therefore set $k_1 = k_0$ and $k_2 = - k_0$,
where $k_0$ is determined by the condition that the single-particle group velocity 
$v_k = dE_k/dk$ is maximum as a function of $k$. The phase shift is then given by
\beq e^{i \theta} ~=~ - ~\frac{1 ~+~ \frac{2 J_z}{J_x}
~e^{- i k_0}}{1 ~+~ \frac{2 J_z}{J_x} ~e^{i k_0}}. \tag{S22} \eeq

\section*{E: Absence of unscrambling in $XXZ$ model and comparison with $XX$ model,
both in a transverse field}

\begin{figure}[H]
\centering
\includegraphics[width=0.6\textwidth]{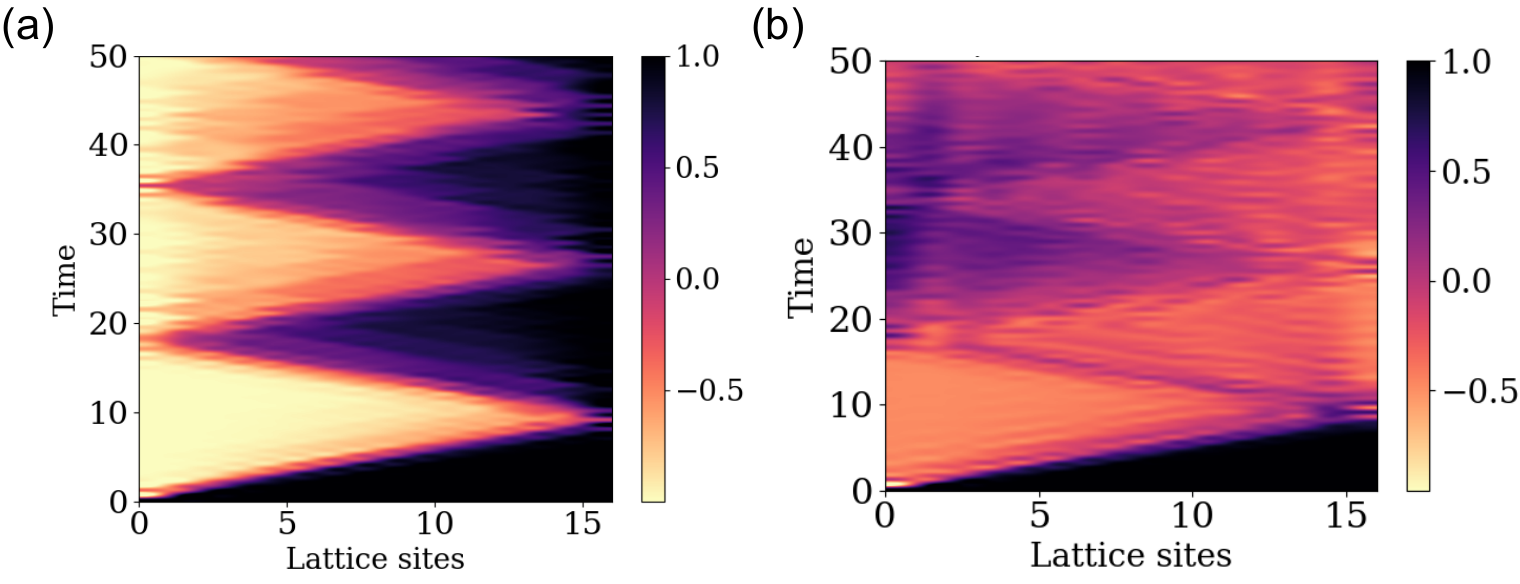}
\caption{$F^{xx}(l,t)$ showing (a) scrambling and unscrambling for the $XX$ spin chain with parameters $J_x = J_y = 1,~ h =8,~ J_z = 0$, and (b) absence of unscrambling in the presence of interactions with the parameters $J_x = J_y = 1,~ h =8,~ J_z = 0.25 $.}
\label{fig:supplement_XXZ} \end{figure}

We present numerical results for $F^{xx}(l,t)$ for the non-interacting $XX$ spin-1/2 chain and the interacting but integrable $XXZ$ chain with a transverse field.
Similar to the TFIM and the interacting non-integrable model discussed in the main text, this pair of models also show scrambling and unscrambling without interactions and absence of unscrambling with interactions. The plots shown in Fig.~\ref{fig:supplement_XXZ} correspond to the parameters, $(a)$  $J_x = J_y = 1,~ h =8,~ J_z = 0$, and $(b)$ $J_x = J_y = 1,~ h =8,~ J_z = 0.25 $. We note that even such a small value of $J_z$ can completely destroy unscrambling as in the case discussed in the main text. This confirms that the absence of unscrambling is not due to non-integrability and can be simply explained by the scattering phase shift argument given in the main text.

\end{widetext}

\end{document}